# A ferrotoroidic candidate with well-separated spin chains

Jun Zhang†[1], Xiancheng Wang†*[1,2], Long Zhou[1], Gangxiu Liu[1,2], Devashibhai T. Adroja[3,4], Ivan da Silva[3],

Franz Demmel[3], Dmitry Khalyavin[3], Jhuma Sannigrahi[3], Hari S. Nair[5], Lei Duan[1,2], Jianfa Zhao[1,2], Zeng Deng[1],

Runze Yu[1], Xi Shen[1], Richeng Yu[1,2], Hui Zhao[1,2], Jimin Zhao[1,2,6], Youwen Long[1,2,6], Zhiwei Hu[7], Hong-Ji Lin[8],

Ting-Shan Chan[8], Chien-Te Chen[8], Wei Wu*[9] and Changqing Jin*[1,2,6]

[1]*Beijing National Laboratory for Condensed Matter Physics, Institute of Physics, Chinese Academy of Sciences, Beijing 100190, China.*

[2]*School of Physical Sciences, University of Chinese Academy of Sciences, Beijing 100190, China.*

[3]*ISIS Facility, STFC, Rutherford Appleton Laboratory, Chilton, Oxford, OX11 0QX, UK.*

[4] *Highly Correlated Matter Research Group, Physics Department, University of Johannesburg, P.O. Box 524, Auckland Park 2006, South Africa.*

[5]*Department of Physics, 500 W. University Ave, University of Texas at El Paso, TX 79968, USA.*

[6] *Songshan Lake Materials Laboratory, Dongguan, Guangdong 523808, China.*

[7]*Max Plank Institute for Chemical Physics of Solids, Nöthnitzer Str. 40, D-01187 Dresden, Germany.*

[8]*National Synchrotron Radiation Research Center (NSRRC), 101 Hsin-Ann Road, Hsinchu 30076, Taiwan.*

[9]*Department of Physics and Astronomy and London Centre for Nanotechnology, University College London, Gower Street, London, WC1E 6BT, UK.*

†These authors contributed equally to this work.

Corresponding Authors: wangxiancheng@iphy.ac.cn; wei.wu@ucl.ac.uk; jin@iphy.ac.cn

The search of novel quasi one-dimensional (1D) materials is one of the important aspects in the field of material science. Toroidal moment, the order parameter of ferrotoroidic order, can be generated by a head-to-tail configuration of magnetic moment. It has been theoretically proposed that one-dimensional (1D) dimerized and antiferromagnetic-like spin chain hosts ferrotoroidicity and has the toroidal moment composed of only two antiparallel spins. Here, we report a ferrotoroidic candidate of $Ba_6Cr_2S_{10}$ with such a theoretical model of spin chain. The structure consists of unique dimerized face-sharing $CrS_6$ octahedral chains along the $c$ axis. An antiferromagnetic-like ordering at ~10 K breaks both space- and time-reversal symmetries and the magnetic point group of $mm'2'$ allows three ferroic orders in $Ba_6Cr_2S_{10}$: (anti)ferromagnetic, ferroelectric and ferrotoroidic orders. Our investigation reveals that $Ba_6Cr_2S_{10}$ is a rare ferrotoroidic candidate with quasi 1D spin chain, which can be considered as a starting point for the further exploration of the physics and applications of ferrotoroidicity.

## Introduction

Ferrotoroidic order, which is one of the four primary ferric order forms, violates both space- and time- reversal symmetries and thus favors magnetoelectric response[1-3]. As the order parameter of ferrotoroidic order, toroidal moment can be generated by a head-to-tail configuration of magnetic moment with $\overrightarrow{T} = \frac{1}{2}\sum_i \vec{r}_i \times \overrightarrow{M}_i$, where $\overrightarrow{M}_i$ is the local magnetic moment at the position $\vec{r}_i$. The classical toroidal moment can be realized by the currents flowing on the surface of a torus along its meridians[4], as seen in Fig. 1(a). Also, it can generally be observed in the single-molecule-based compounds with unique architectures of wheel-shaped topology[5], such as $Dy_6$ wheels[6, 7], $Dy_4$ squares[8] and $Dy_3$ triangles[9], which are sketched in Fig. 1(b-d), respectively. In the system of crystalline solids, the spontaneous toroidization of toroidal moment, that is the ferrotoroidic order, has drawn increasing attentions due to its novel asymmetry properties and potential applications[2-5, 10-15]. Several ferrotoroidic candidates have been proposed[3, 15], such as the orthophosphates of $LiCoPO_4$ with olivine structure[10] and the pyroxene structure type of $LiFeSi_2O_6$[16]. $LiCoPO_4$ is the prominent example in which hallmark evidences have been presented for its ferrotoroidicity. In $LiCoPO_4$, the toroidal domains have been detected by optical second harmonic generation and shown to be switched by a conjugate toroidal field with hysteretic behavior[10, 17]. In the spin ordering state of $LiCoPO_4$[18], two pairs of $Co^{2+}$ spins, which are located at the positions like (1/4+ε, 1/4, -δ) in the unit cell with the small displacements of ε and δ allowed by the symmetry, form opposite but unequal toroidal moments, and thus cause a net $\overrightarrow{T}$. In addition, $LiFeSi_2O_6$ exhibits polarization in all the off-diagonal tensor components, which suggests the primary order parameter should be the toroidal moment[16]. Besides the experimentally proposed ferrotoroidic candidates, it is theoretically predicted that an dimerized and antiferromagnetic-like (AFM-like) spin chain, as shown in Fig 1(g), should host ferrotoroidicity[15], where the distance of spins $a_1 \neq a_2$. In the model chain both the space- and time-inversion symmetries are broken due to the "spin-pairing" and the toroidal moment consists of only two antiparallel spins. However, such a ferrotoroidic candidate with quasi 1D structure has not been experimentally reported so far.

To date, most of the ferrotoroidic candidates have been reported to be oxides with high-dimensional structure. Quasi 1D structures are observed more frequently in the chalcogenides because their chemical bonds are in general less ionic than those in oxides[19, 20]. To look for such a 1D ferrotoroidic chain in quasi 1D system, we turn to the $Hf_5Sn_3Cu$-anti type ternary compounds $A_3BX_5$ (A denotes the alkali earth metal, B the transition metal and X the chalcogen), which have been intensively studied due to their strong 1D structural characteristic[21-25]. In present work, we successfully synthesized the new compound $Ba_6Cr_2S_{10}$ under high-pressure and high-temperature conditions. It possesses dimerized face-sharing $CrS_6$ octahedral chains that are separated by 9.1228 Å. $Ba_6Cr_2S_{10}$ undergoes an AFM-like transition at ~10 K. Due to the dimerized characteristic, the AFM-like spin chains in $Ba_6Cr_2S_{10}$ are exactly the theoretically proposed model of spin chains. In addition, the magnetic point group is *mm'2'*, which allows three ferroic orders in $Ba_6Cr_2S_{10}$: (anti)ferromagnetic, ferroelectric and ferrotoroidic orders[3]. Therefore, $Ba_6Cr_2S_{10}$ is considered as the rare 1D ferrotoroidic candidate.

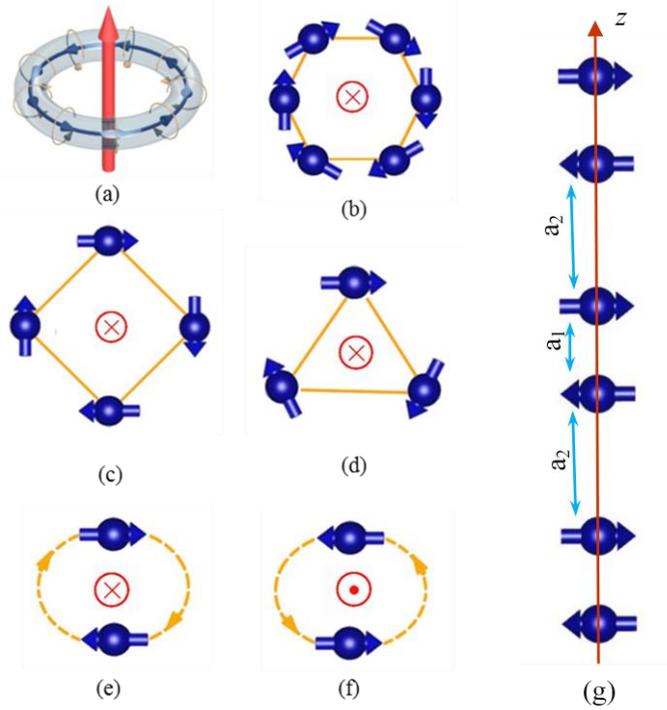

Fig. 1 The magnetic toroidal moments generated by (a) the currents flowing on the surface of a torus along its meridians[4]; (b-d) six, four and three head-to-tail configurations of magnetic moments, respectively; (e, f) antiparallel spins with equal but opposite toroidal moments. (g) dimerized and AFM-like spin chain, in which the magnetic ions are paired with spin space $a_1 \neq a_2$.

## Results and Discussions

**Crystal structure.** Polycrystalline samples of $Ba_6Cr_2S_{10}$ have been synthesized under high-pressure and high-temperature conditions, from which the needle-like single crystals with the size of about 100 μm can be selected. The chemical composition measured on the single crystals, as shown in Fig. S1, presents an average atomic ratio of Ba : Cr : S = 3.1(2):1.0(0):4.9(3), which is close to the expected stoichiometry of $Ba_6Cr_2S_{10}$.

Table 1 The summary of the crystallographic data from the refinement of powder X-ray diffraction measured at room temperature for $Ba_6Cr_2S_{10}$

| Formula: | $Ba_6Cr_2S_{10}$ | | | | |
|---|---|---|---|---|---|
| Space group: $P\text{-}62c$ (*No. 190*); Temperature/K: 300; | | | | | |
| $a=b=9.1228(3)$ Å , $c=12.3643(2)$ Å ; $\alpha=\beta=90°$, $\gamma=120°$; | | | | | |
| $Z=2$; V= 891.161(5) Å$^3$; | | | | | |
| Final $R$ indexes [all data]: $\chi^2=3.686$, $R$p = 2.68%, wRp = 4.25% | | | | | |
| Atomic parameters | | | | | |
| Atom | wyck. | $x$ | $y$ | $z$ | U(eq) |
| Ba1 | 6g | 0.6175(1) | 0 | 0 | 0.012 |
| Ba2 | 6h | 0.3926(7) | 0.0195(2) | 0.25 | 0.004 |
| Cr1 | 4e | 0 | 0 | 0.1206(5) | 0.013 |
| S1 | 4f | 0.3333 | 0.6666 | 0.1521(2) | 0.011 |
| S2 | 6g | 0.2083(4) | 0 | 0 | 0.016 |
| S3 | 6h | 0.2173(4) | 0.2212(8) | 0.25 | 0.008 |
| S4 | 4f | 0.3333 | 0.6666 | 0.6302(1) | 0.018 |

The crystal structure was solved by the combination of single crystal and powder XRD studies. The structural model with the highest symmetry of space group that can satisfactorily describe both the single crystal and powder XRD data is a hexagonal structure with $P\text{-}62c$. Using the positional parameters obtained by the single crystal XRD, a Rietveld refinement of the powder XRD data was carried out as shown in Fig. 2(a). The refinement was smoothly converged to $\chi^2 = 3.686$, $R$p = 2.68% and w$R$p = 4.25%. Table 1 shows the summary of the partial crystallographic data. The lattice constants $a = b = 9.1228(3)$ Å and $c = 12.3643(2)$ Å.

The crystal structure of $Ba_6Cr_2S_{10}$ is sketched in Fig. 2(b-d). It consists of face-sharing $CrS_6$ octahedral chains along the $c$ axis, which are arranged in the form of triangular lattice in the $ab$-plane and separated by Ba and S ions. In the unit cell, all the Cr sites are equivalent, such as (0, 0, 1/8-ε), (0, 0, 3/8+ε), (0, 0, 5/8-ε) and (0, 0, 7/8+ε), where ε=0.00435 is the small displacement allowed by the symmetry. The in-chain distance of the adjacent Cr atom can be calculated with the formula of $d=(2/8\pm2ε)*c$, here $c$ is the lattice constant of $c$ axis, which leads to two different distances ($d_{\text{in-chain}}$) of 3.1986(5) Å and 2.9835(4) Å. That is the Cr ions within the spin chain are dimerized. However, the distance ($d_{\text{in-plane}}$) between the adjacent Cr ions in the $ab$-plane is 9.1228(3) Å, which is significantly larger than $d_{\text{in-chain}}$ and thus, demonstrates a characteristic of quasi-1D spin chains in the view of crystal structure. In fact, the compound $Ba_3CrS_5$ has been reported already. However, it was proposed to crystallize in a similar crystal structure with $P6_3cm$ space group, where the lattice constants *a* and *b* are approximately the same as those of $Ba_6Cr_2S_{10}$, but the *c* axis is the half[26]. The double *c* axis and the alternately spaced Cr atoms in the *c* axis reveal that the $CrS_6$ chains in our sample of $Ba_6Cr_2S_{10}$ is dimerized.

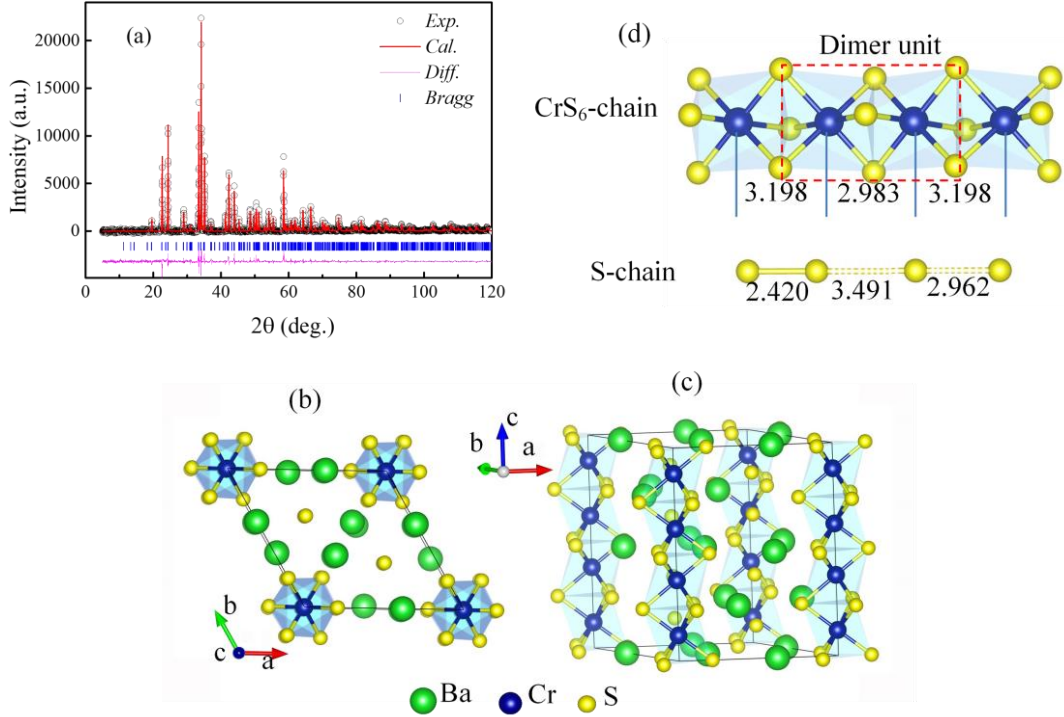

Fig. 2 (a) The XRD pattern and the refinement for $Ba_6Cr_2S_{10}$. (b-d) The schematic map of the crystal structure of $Ba_6Cr_2S_{10}$; (b) the top view with the projection along [001] direction and (c) the side view showing a chain structure; (d) the sketch of the dimerized octahedron $CrS_6$ chain and S-chain.

**Dimerized character.** To further confirm the dimerized structure, the STEM experiments were performed on polycrystalline sample of $Ba_6Cr_2S_{10}$. Fig. 3(a) shows the high-angle annular dark-field (HAADF) image along the [100] zone axis, where the bright dots indicate the Ba atomic columns. In the HAADF image, the Ba ions are aligned periodically with a zigzag arrangement along *c* axis, as marked by alternating red and blue lines. Also, Fig. 3(a) displays the schematic map of the structure with the projection along the [100] zone axis for $Ba_6Cr_2S_{10}$ with dimerized $P$-62$c$ space group, where the arrangement of Ba ions is zigzag-like and is the same as the HAADF image. While for a structure with non-dimerized $P6_3cm$ space group, the Ba ions should be straightly aligned along *c* axis. Therefore, the HAADF image presents a good consistence with that of the dimerized structure solved from our XRD data.

Besides the $CrS_6$ chains, S-chains in the structure are located at the center of the triangular lattice as shown in Fig. 2(b). It is noted that the distances between the adjacent S ions in the S-chains are 2.420 Å, 3.491 Å to 2.962 Å, respectively (Fig. 2(d)). The small distance of 2.420 Å in the S-chains is very close to the bonded length of S ions, suggesting the formation of $S_2^{2-}$. Thus the valence state of S ions in the $S_2^{2-}$ pairs should be -1. On the contrary, for the non-dimerized structure, the S-S covalent bond in the S-chains should be absent because the S ions are aligned with equal distance of 3.097 Å in the S-chain[26]. To further investigate the oxidation state of the S ions, we collected the sulfur $K$-edge absorption spectra using $Cr_2S_3$ as the reference standard, as shown in Fig. 3(b). For $Ba_6Cr_2S_{10}$, there are two peaks for sulfur $K$-edge absorption centering at 2468.5 eV and 2469.8 eV corresponding to S-S bonding and Cr-S bonding, respectively, which confirms the formation of $S_2^{2-}$.

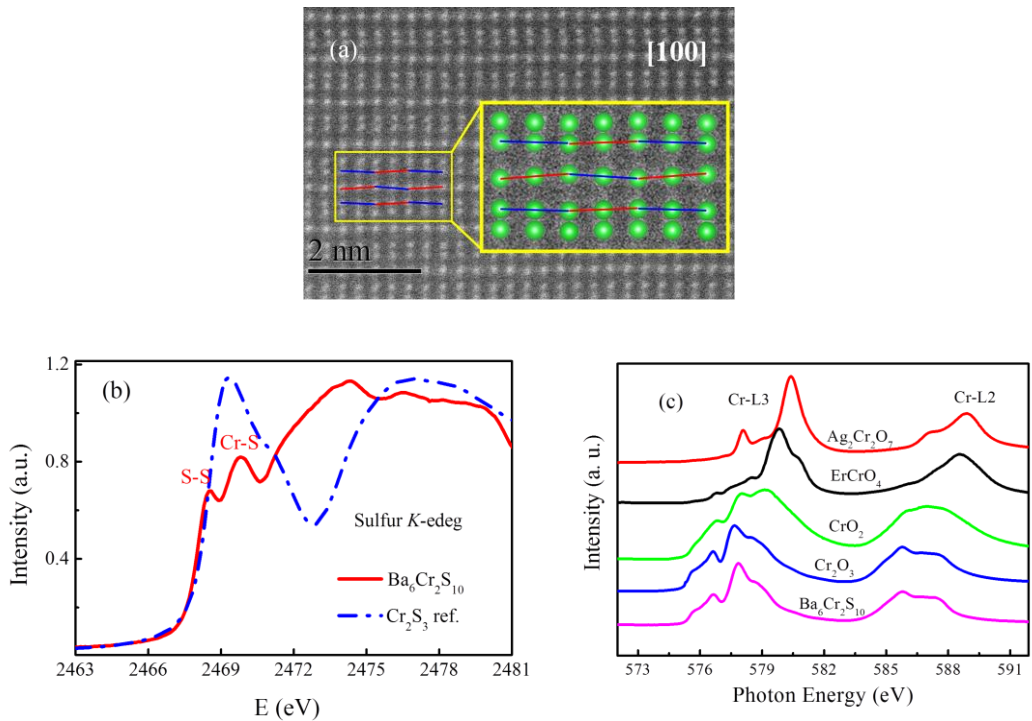

Fig. 3　(a) The HAADF image along the [100] zone axis and the sketch maps of dimerized structure ($P$-62$c$) of $Ba_6Cr_2S_{10}$ with the projection along $a$ axis. The large green circles denote Ba ions. The red and blue dotted lines indicate a zigzag arrangement of Ba ions. (b) The sulfur K-edge XAS measurements for $Ba_6Cr_2S_{10}$ sample with $Cr_2S_3$ as the reference material. (c) The Cr $L$-edge XAS spectra of $Ba_6Cr_2S_{10}$ together with $Cr_2O_3$, $CrO_2$, $ErCrO_4$, $Ag_2Cr_2O_7$ as a $Cr^{3+}$, $Cr^{4+}$, $Cr^{5+}$ and $Cr^{6+}$ reference materials, respectively.

Formally one would expect $Cr^{4+}$ in $Ba_6Cr_2S_{10}$ in case of $S^{2-}$ state. Thus, a formation of S-S covalent bond can be further confirmed by real valence state of Cr ion using the soft XAS at Cr $L_{2,3}$ edge, which is highly sensitive to the valence state and local environment of 3$d$ transition metal[27-31]. As seen from Fig. 3(c) and Fig. S2 (a-c), the spectra weight systematically shift to higher energy from $Cr_2O_3$ to $CrO_2$, and $ErCrO_4$ further to $Ag_2Cr_2O_7$ from bottom to top. The similar multiplet spectral feature and the energy position at the of Cr $L$-edge of $Ba_6Cr_2S_{10}$ and $Cr_2O_3$ in Fig. 3(c) suggest a $Cr^{3+}$ valence state in $Ba_6Cr_2S_{10}$, therefore firmly confirming the formation of S-S bond. Both the sulfur *K*-edge and Cr *L_{2,3}* edge XAS experiments reveal the formation of S-S covalent bond and hint the dimerized structure of $Ba_6Cr_2S_{10}$. We note that the dimerized structure should be retained at low temperature because no abnormal corresponding structural transition has been observed in the specific heat data, as shown in Fig. S3. Based on the site symmetry and the analysis of valence state, the molecular formula of $Ba_6Cr_2S_{10}$ can be rewritten as $Ba^{2+}_6Cr^{3+}_2S^{2-}_6(S^{2-}_2 S^{1-}_2)$, where the charge balance is reached.

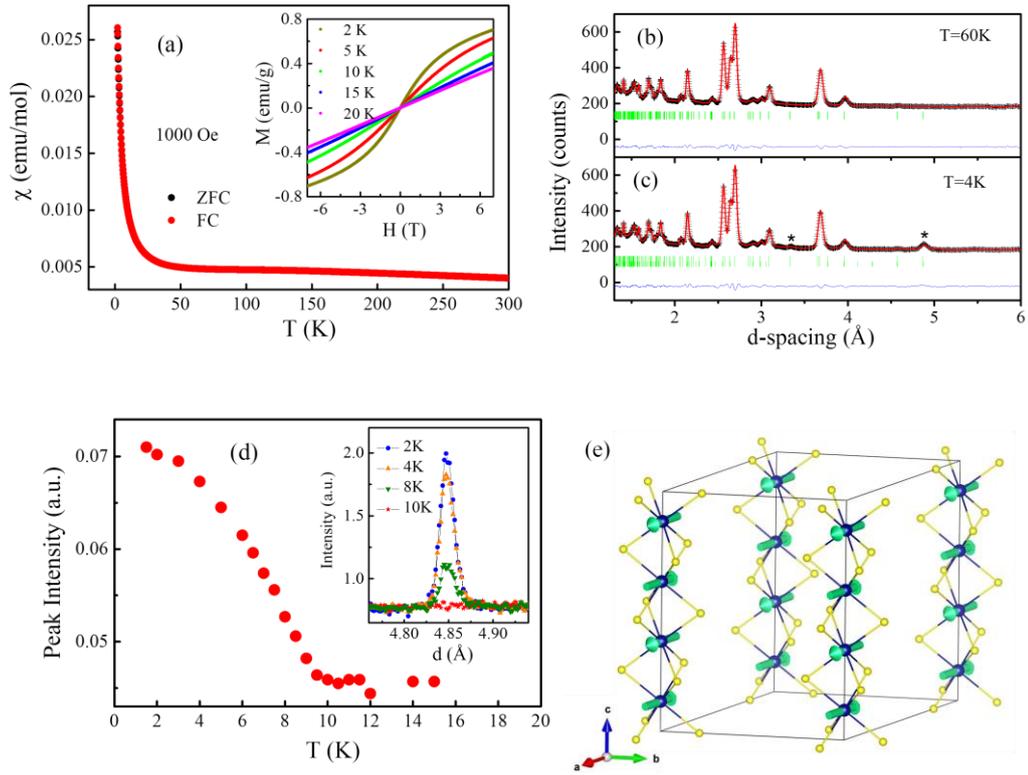

Fig. 4 (a) The temperature dependence of magnetic susceptibility χ measured with magnetic field 1000 Oe. The inset is the M-H curves measured at different temperatures. (b,c) The Rietveld refinement of the neutron diffraction pattern collected at T=60 K ($R_{Bragg}$=3.6%) and T=4 K ($R_{Bragg}$=2.9%, $R_{magnetic}$=5.6%), respectively. The magnetic intensities appearing below ordered temperature are marked with asterisks. (d) The magnetic Bragg peak intensity as a function of temperature. (e) The magnetic structure of $Ba_6Cr_2S_{10}$ with the *Ama'2'* magnetic space group, which was directly determined from the neutron diffraction data with the ordered moment ~1.03(9) μ_B.

**Magnetic properties and magnetic structure.** Fig. 4(a) displays the temperature dependence of *dc* magnetic susceptibility. The susceptibility increases sharply at low temperature, which presents a slightly canted AFM-like phase transition behavior. The inset is the M-H curves measured at different temperatures. At 2 K, the magnetization shows a ferromagnetic (FM) behavior, evidencing the presence of FM component in this system. When increasing temperature to 10 K, the M(H) curve approaches a straight line, suggesting a canted AFM-like transition occurs at ~10 K. The inverse of susceptibility as a function of temperature is shown in Fig. S4. The $1/\chi(T)$ in high temperature region follows the Curie-Weiss law. By using the formula $\chi^{-1} = (T - \theta_p)/C$, the data can be well linearly fitted, and the effective magnetic moment is estimated to be $\mu_{eff} = 4.04 \ \mu_B/Cr$ and the Curie-Weiss temperature $\theta_p = -714$ K. The value of $\mu_{eff}$ is close to that expected for $Cr^{3+}$ ions with S=3/2, and agrees with the results of Cr L-edge XAS experiments. The negative sign of $\theta_p$ suggest an AFM exchange interaction between the Cr spins within the chains.

To examine the spin arrangement or magnetic structure of $Ba_6Cr_2S_{10}$, the neutron powder diffraction (NPD) was performed on $Ba_6Cr_2S_{10}$ and the results are shown in Fig. 4(b-e). Below $T_N$, the NPD data provide the evidence of several additional magnetic Bragg peaks at lower moment transferred Q, which can be indexed with magnetic propagation vector $\boldsymbol{k}$=(0,0,0), as shown in Fig. 4(b, c). Fig. 4(d) presents the magnetic Bragg peak intensity as a function of temperature, which demonstrates clearly that the long-range spin ordering transition temperature is ~10 K. To solve the magnetic structure, we adapted a symmetry-based approach implemented into ISOTROPY and ISODISTORT software. There are four one-dimensional ($\Gamma_i$ i=1-4) and two two-dimensional ($\Gamma_i$, i=5-6) irreducible representations (IR) associated with magnetic propagation vector $\boldsymbol{k}$=(0,0,0) and entering into the magnetic reducible representation on the 4e Wyckoff position of Cr. $\Gamma_1$- $\Gamma_4$ have one basis vector for each of them. $\Gamma_3$ gives FM moments along the c-axis, while $\Gamma_1$, $\Gamma_2$, and $\Gamma_4$ give FM coupling in the ab-planes and AFM coupling along the c-axis with spin configurations, '-++-', '-+-+' and '+--+', respectively. $\Gamma_5$ and $\Gamma_6$ have four basis vectors for each, indicating an ordered state with magnetic moments FM in the ab-plane and with different AFM coupling along the c-axis. All of these IRs were tested in the refinement procedure against the experimental data. The important experimental observation, coming from the magnetization measurements (Fig. 4(a)), is the presence of a spontaneous FM component (a canted AFM). The only solution which provides a good refinement quality and is compatible with the FM component implies a magnetic Ama'2' symmetry. To a good approximation the magnetic structure is collinear and involves AFM coupling along the chain between the nearest-neighboring spins, which are aligned along the a or b axis of the parent hexagonal P-62c space group, as shown in Fig. 4(e). The estimated $Cr^{3+}$ ordered state moment is 1.03(9) $\mu_B$, which is much smaller than the fully ordered moment ($3\mu_B$) expected from the susceptibility measurements. Since strong quantum fluctuation generally exists in low dimensional system and prevent long-range ordering formation, the reduction of the ordered moment is a common phenomenon in the system with quasi 1D spin chains[32-34]. This symmetry allows two additional orthogonal modes as shown in Fig. S5(b, f), an AFM chain with spin in the *ab*-plane and a FM chain with spin along the *c*-axis. These modes, however, are too small to be detected directly from our neutron diffraction data and the refinement of the data is unstable or resulted in values which were not statistically significant. Although Cr cations in $Ba_6Cr_2S_{10}$ take a single Wyckoff position in their paramagnetic space group, the proposed magnetic structure with the Ama'2' space group allows coupling of orthogonal antiferromagnetic and ferromagnetic modes. That is, in this case, Dzyaloshinskii-Moria (DM) antisymmetric exchange can be activated and lead to the canting and couple ferromagnetic component, as observed in magnetization experiments. An interesting observation is that the magnetic symmetry is polar (magnetic point group is *mm'2'*) and the polar axis is perpendicular to the spontaneous magnetization. Thus, this compound is a rare example where the magnetic ordering combines all three orthogonal ferroic order parameters; polarization P, magnetization M and toroidal moment $\vec{T} = \frac{1}{2}\sum_i \vec{r}_i \times \vec{M}_i$. In addition, the energies of different possible magnetic ground states (Seen in Fig. S5(a-f)) were calculated. The calculation results show that the configuration of ( ↑ ↓ ↑ ↓ ) with spin oriented along the *a* axis has the lowest energy, which agrees with the experimental results.

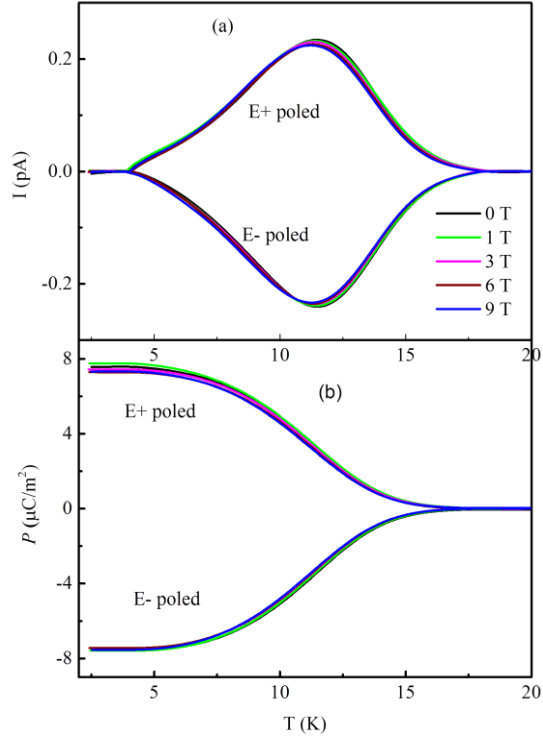

Fig. 5 For the polycrystalline sample of $Ba_6Cr_2S_{10}$, (a) the pyroelectric current $I_p$ poled from 2 K to 20 K and (b) the electric polarization $P$ measured under 0 T, 1 T, 3 T, 6 T and 9 T, respectively.

**Pyroelectric current and polarization.** Since the magnetic symmetry is polar, polarization should be expected under the magnetic state for the $Ba_6Cr_2S_{10}$ sample. Fig. 5(a, b) shows the pyroelectric current $I_p$ poled from 20 K to 2 K and the electric polarization $P$ for the polycrystalline sample of $Ba_6Cr_2S_{10}$. There is a peak centered at ~11.5 K in the $I_p(T)$ curve, which is close to the long-range spin ordering (LRSO) temperature. The magnitude of the electric polarization is about 7.8 $\mu C/m^2$. In addition, the sign of $I_p$ and $P$ can be inverted symmetrically by changing the sign of poled electric field. When the applied magnetic field increases from 0 T to 9 T, the value of $P$ is slightly changed while the $I_p(T)$ peak shifts towards low temperature, as can be clearly seen in Fig. S6(a-c). For $Ba_6Cr_2S_{10}$, the magnetic point group is $mm'2'$, which allows the occurrence of polarization[3]. The observation of polarization near the LRSO ordering temperature reveals that the polarization and the spin ordering are coupled each other, and the shift of $I_p(T)$ peak suggests the external magnetic field should be against AFM coupling spin order. The $I_p(T)$ peak denotes the temperature where polarized charges have the fastest decrease with increasing temperature. The unreleased polarized charges above the LRSO (~10 K) might arise from the existence of small polarized domains related with intrachain spin short-range correlation.

**Discussions.** $Ba_6Cr_2S_{10}$ possesses a strong 1D characteristic. In the view of structure, the $CrS_6$ spin chains are >9 Å apart, suggesting a very weak interchain coupling strength. In addition, no abnormality was observed in the specific heat data corresponding to the magnetic transition of ~10 K, as shown in Fig. S3, which demonstrates a strong 1D property. For an ideal spin chain system, LRSO cannot occur in the finite temperature because of strong thermal or quantum fluctuation[35]. In a quasi 1D spin compound, although the inter-chain exchange interaction is much weaker than that in the intra-chain, it governs the long-range magnetic transition[36]. Before the formation of LRSO, the intra-chain interaction has led to the gradually development of short-range order within the chains and thus, the magnetic entropy has been released far above the LRSO transition. As a result, the abnormality associated with the magnetic transition is usually small or even invisible in the specific heat data in quasi 1D system. Here, for our sample, no abnormality in the heat specific data related with the magnetic transition reveals the strong 1D characteristic of $Ba_6Cr_2S_{10}$.

The studies of the crystal structure reveal that $Ba_6Cr_2S_{10}$ is composed of well-defined quasi-1D dimerized $CrS_6$ octahedral chains. The magnetic measurements further prove that below the ordered temperature the spins are oriented along the *a* (or *b*) axis, AFM coupled along the chain direction and FM coupled in the *ab*-plane. Thus, $Ba_6Cr_2S_{10}$ consists of dimerized and AFM-like spin chains as shown in Fig. 6(b), where the distance between neighboring Cr ions in the chain is $a-2\delta = 2.9835(4)$ Å and $a+2\delta =3.1986(4)$ Å, respectively. Here, $a=3.0910(4)$ Å is the distance of neighboring Cr ions in uniform AFM chain as seen in Fig. 6(a), and $\delta$=0.0538(4) Å is the displacement of Cr ion. The dimerized and AFM-like spin chain contains the toroidal moment with just two antiparallel spins, breaks both space- and time-reversal symmetries, and is proposed to be a simple ferrotoroidic model. We note that the dimerized structure is necessary for the AFM-like chain to host ferrotoroidictiy. As shown in Fig. 6(a), the uniform AFM spin chain is non-toroidal. First, the system cannot host a macroscopic toroidal moment because it keeps space-reversal symmetry according to the spin arrangement and each spin site. The second, in fact, the uniform AFM spin chain also is time-reversal symmetric. Although it breaks time-reversal symmetry microscopically under the time inversion transformation, it can go back by translating all the spins by half a unit length of $a$ along the $z$ axis. Therefore, time-reversal symmetry is not broken macroscopically in the uniform AFM spin chain. According to the study by Ederer[15], for the uniform AFM spin chain, the allowed toroidization values are centrosymmetric with $\vec{T}_n = -(\frac{1}{2} + n)\frac{S}{2}\hat{y}$, where $n$ is any integer number, S is the moment on the spin site. Analogous to the case of electric polarization, the toroidization values allowed by centrosymmetry means a nontoroidal state. So, the uniform AFM spin chain does not host ferrotoroidictiy. Here, ignoring the canting feature, we can estimate the spontaneous toroidization $\vec{T}_s$ for a single $CrS_6$ chain by the difference between its ferrotoroidic state and nontoroidic state, as shown in Fig. 6. That is $\vec{T}_s = \frac{1}{2}\frac{2\delta}{\Omega_1}S\hat{y} = 8.9 \times 10^{-3}\mu_B/Å^2\hat{y}$, where $S$ =1.03(9) $\mu_B$ is the measured ordering moment of $Cr^{3+}$ ion, $\delta$=0.0538(4) Å is the displacement of Cr ions and $\Omega_l = 2a$=6.1821 (4) $Å^3$ is the 1D unit cell. The positions $\vec{r}_i$ and the magnetic moment $\vec{M}_i$ of Cr cations in $Ba_6Cr_2S_{10}$ were shown in Table S1. For $Ba_6Cr_2S_{10}$, the spontaneous toroidization is given by $\vec{T}_s = -\frac{1}{2}\frac{16\delta}{\Omega_2}S\hat{y} = 4.9 \times 10^{-4}\mu_B/Å^2\hat{y}$, where the unit cell volume of $Ba_6Cr_2S_{10}$ is $\Omega_2$ =891.161(5) $Å^3$.

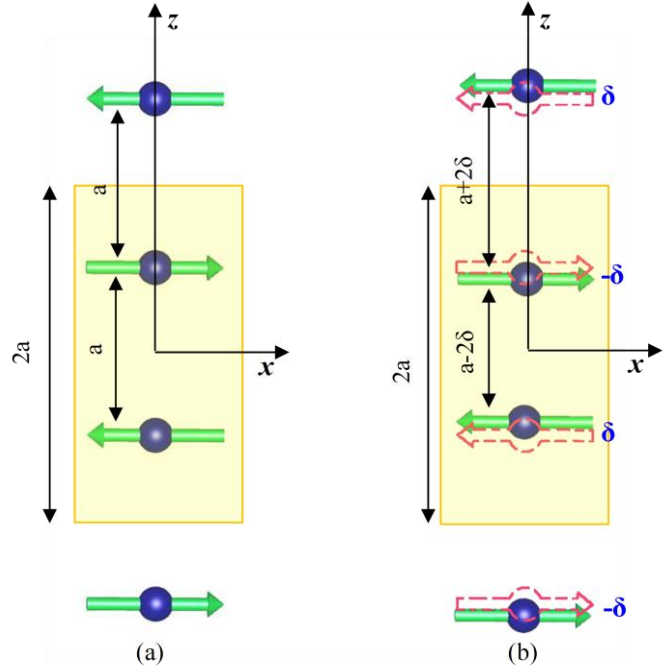

Fig. 6 (a) The sketch of the uniform AFM spin chain with the distance between neighboring spins $a$, which is a non-toroidal state; (b) the dimerized and AFM-like spin chain with the distance of $a$-$2\delta$ = 2.9835(4) Å and $a$+$2\delta$ =3.1986(4) Å, respectively, which hosts the simple ferrotoroidicity structure. The spin chain with toroidal state can be obtained from the chain with non-toroidal state through shifting the spins along $z$ axis by the displacement of $\pm\delta$.

As the fourth primary ferric order form, ferrotoroidicity violates both space- and time-reversal symmetries, which favors magnetoelectric response and thus provides potential pathways for applications such as in information storage. To our knowledge, $Ba_6Cr_2S_{10}$ is the rare example that consists of dimerized and AFM-like spin chains, which can be considered as a starting point for further exploration of the magnetoelectric coupling from the point of view of both fundamental theory and potential application. In addition, because of the dimerized characteristic in $Ba_6Cr_2S_{10}$ structure, it is indeed a system with alternating-exchange AFM-like spin chain. For this chain system with alternated AFM coupling strength, the ground state is predicted to be non-magnetic spin singlet state[37, 38], which is separated from the excited triplet state by an energy gap. This energy gap can be suppressed to zero by external magnetic field or inter-chain exchange interaction, where Bose-Einstein condensation occurs and a long-range magnetic order transition is induced that represents a quantum critical point. The exotic phenomenon has been observed in the system with spin ladders[39-43]. Therefore, the long-range magnetic order in $Ba_6Cr_2S_{10}$ should be induced by inter-chain coupling and the collective excitation of spin, namely the behavior of magnon should be interesting for further studies.

## Conclusion

In summary, we discovered a new material $Ba_6Cr_2S_{10}$ under high pressure conditions. The crystal structure consists of dimerized $CrS_6$ octahedral chains, exhibiting strong 1D structural character. The spin order forms at ~10 K and the spins are confined within the *ab*-plane, coupled antiferromagnetically along the chain with the estimated ordered state moment 1.03(9) $\mu_B$ for each $Cr^{3+}$. Our results reveal that, in the spin chains of $Ba_6Cr_2S_{10}$, the spins are alternately spaced and antiparallelly aligned and thus, $Ba_6Cr_2S_{10}$ is a rare quasi-1D ferrotoroidic example, where the toroidal moment is only composed of two antiparallel spins. What's more, the magnetic point group of *mm'2'* allows three ferroic orders of AFM, ferroelectric and ferrotoroidic, which offers an opportunity to further study the properties of ferrotoroidicity.

## Experimental Sections

**Sample synthesis.** $Ba_6Cr_2S_{10}$ samples were synthesized under high pressure and high temperature conditions. The starting materials are commercially available crystalline powders of Cr (Alfa, >99.9% pure), S (Alfa, >99.999% pure) and lumps of Ba (Alfa, immersed in oil, >99.2% pure). The precursor BaS was prepared by heating the mixture of Ba blocks and S powder in an alumina crucible sealed in an evacuated quartz tube at 700 $^o$C for 20 h. The obtained BaS powder, Cr and S were homogenously mixed according to a molar ratio 3:1:2. The pellet of the mixture underwent a treatment under the conditions of 5.5 GPa and 1200 $^o$C. After the high pressure and high temperature process, the black pure polycrystalline sample of $Ba_6Cr_2S_{10}$ was obtained, from which the single crystals with the size of 100 micrometers can be selected.

**Structure characterization.** The chemical composition of the $Ba_6Cr_2S_{10}$ single crystal was determined by energy dispersive X-ray spectroscopy (EDX). The aberration-corrected scanning transmission electron microscopy (STEM) was performed on a JEM-ARM200F microscope with double Cs correctors for the condenser lens and objective lens. The available spatial resolution for each of the STEM images is better than 78 pm at 200 kV. Single crystal X-ray diffraction (XRD) measurements were performed at room temperature on an APEX III CCD diffractometer using monochromatic Mo Kα radiation. The crystal structure was solved by the Patterson methods and refined by full-matrix least-squares fitting on $F^2$ within the Olex2 crystallographic software package. The structure was checked with PLATON for missing symmetry elements. Powder XRD measurements were carried out on a Rigaku Ultima VI (3KW) diffractometer using Cu Kαradiation generated at 40 kV and 40 mA. The XRD data were collected with a scanning rate of 1 $^{\circ}$ per minute and a scanning step length of 0.02 $^{\circ}$. The Rietveld mode in the GSAS software packages was used for the refinement of the diffraction spectra. Soft x-ray absorption scattering (XAS) at the Cr $L_{2,3}$ edges and the sulfur $K$-edge were measured at the beamline BL11A and BL16A, respectively, of the National Synchrotron Radiation Research Center (NSRRC) in Taiwan.

**Magnet and transport characterization.** Neutron powder diffraction (NPD) measurements were carried out using the time-of-flight (TOF) General Materials (GEM) diffractometer and the OSIRIS diffractometer at the ISIS Facility of the Rutherford Appleton Laboratory (UK) [44]. The NPD patterns were refined by using a multipattern Rietveld mode in FullProf program. The resistivity was measured using a physical property measurement system (PPMS), and the dc magnetic susceptibility was measured using a superconducting quantum interference device (SQUID).

**Polarization characterization.** The pyroelectric current $I_p$ was measured on the sample with the size of Φ3 mm and 0.25 mm in thickness, before which the sample was electrically poled by applying an electric field $E$ of ±6.6 kV at 20 K. The electric polarization $P$ was obtained via the integration of $I_p$ as a function of time.

**Supporting information**

Supporting Information is available from the Wiley Online Library.

Fig. S1-S6, energy dispersive X-ray spectrum, XAS, specific heat, magnetic susceptibility, first-principles calculations and polarization.

Table S1 The sites of Cr cations for $Ba_6Cr_2S_{10}$.

**Acknowledgements**

We are grateful to Prof. Manfred Fiebig and Prof. Nicola Spaldin for the helpful discussion about the ferrotoroidicity. The work was supported by National Key R&D Program of China and the Natural Science Foundation of China (Grant No. 2018YFA0305700, 11974410, 12104488, 2018YFE0103200, 2017YFA0302900, 2017YFA0303603, 11934017, 11921004, 11904392, and 11774408), and Chinese Academy of Sciences (Grant No. XDB33000000). D.T.A. would like to thank the Royal Society of London for UK-China Newton mobility funding and Newton Advance Fellowship funding. We acknowledge support from the Max Planck-POSTECH-Hsinchu Center for Complex Phase Materials.

**Author contributions**

Proposal of the material: X.C.W. and C.Q.J.; research coordination: X.C.W., C.Q.J., J.Z. and W.W.; synthesis, X-ray diffraction and magnetization: J.Z., X.C.W., L.D., Z.D., R.Z.Y., H.Z., J.M.Z., and J.L.Z.; pyroelectric current and polarization: L.Z., G.X.L. and Y.W.L.; X-ray absorption spectroscopy: J.F.Z., Z.W.H, H.J.L. and C.T.C; electron microscopy (STEM):X.S and R.C.Y; neutron diffraction: D.T.A., I.D.S., F.D., D. K., J.S. and H.S.N; theory calculation: W.W.; manuscript: J.Z., X.C.W.,C.Q.J. and W.W..